\documentclass[prb,twocolumn,preprintnumbers,superscriptaddress]{revtex4}

\usepackage{graphicx}
\usepackage{dcolumn}
\usepackage{amsmath}
\usepackage{color}

\begin{document}

\title{Interplay of electronic correlations and chemical bonding in FeN$_2$ under pressure}

\author{I. V. Leonov}
\affiliation{M. N. Mikheev Institute of Metal Physics, Russian Academy of Sciences, 620108 Yekaterinburg, Russia}
\affiliation{Institute of Physics and Technology, Ural Federal University, 620002 Yekaterinburg, Russia}

\begin{abstract}

We report a theoretical study of the effects of electronic correlations, magnetic properties, and chemical bonding in the recently synthesized high-pressure orthorhombic phase of FeN$_2$ using the DFT+dynamical mean-field theory approach. Our analysis documents a complex crystal-chemical behavior of FeN$_2$ characterized by the formation of a strongly covalent N-N bond with an unexpected valence state of Fe ions $3+$ (paramagnetic ferric Fe$^{3+}$ ions in the low-spin state), in agreement with available experimental data. Our results reveal weak (orbital-dependent) correlation effects, which are complicated by the possible emergence of multiple spin density wave states on a microscopic level. This suggests the importance of antiferromagnetic spin fluctuations to explain the properties of FeN$_2$ under pressure.
  
\end{abstract}

\maketitle

\section{Introduction}

Under high-pressure conditions crystalline solids, e.g., transition metal compounds, are known to undergo remarkable transformations associated with alterations of their quantum state 
and the emergence of unusual stoichiometries \cite{McMillan_2006,Miao_2020,Zhang_2017,Oganov_2019,Dubrovinsky_2022}. 
A particular example is the binary Fe$-$O system. For iron oxides the high-pressure and -temperature synthesis gives numerous unexpected compositions with complex crystal structures \cite{Bykova_thesis,Bykova_2016}. Under pressure iron oxides adopt different compositions, such as FeO$_2$, Fe$_2$O$_3$, Fe$_3$O$_4$, Fe$_4$O$_5$, Fe$_5$O$_6$, Fe$_5$O$_7$, Fe$_7$O$_9$, etc., \cite{Bykova_2016,Lavina_2015,Lavina_2016,Sinmyo_2016,Ovsyannikov_2018,Ovsyannikov_2020,
Greenberg_2018,Hikosaka_2019,Leonov_2019,Layek_2022,Greenberg_2023,Hu_2016,Nishi_2017,
Wright_2002,Senn_2012,Perversi_2019,Baldini_2020,Kunes_2009,Ohta_2012,Leonov_2015,
Leonov_2016,Leonov_2020}, with unusual crystal structures and complex electronic and magnetic properties, which can be systematized with a homologous structural series $n\mathrm{FeO} \cdot m \mathrm{Fe}_2\mathrm{O}_3$ (with the exception of FeO$_2$) \cite{Bykova_2016}. 

Overall, this reflects the complexity of crystal chemistry and chemical bonding in these solids, complicated by the effects of strong electron-electron correlations in the partially occupied Fe $3d$ orbitals \cite{Mott_1990, Imada_1998}. In fact, many of these compounds exhibit a strongly correlated metallic or Mott-Hubbard (charge-transfer) insulating behavior at low pressure and temperature.

The binary transition metal-nitrogen systems and, in particular, nitrogen-rich compounds $M$N$_m$ with $m>1$, are another prominent example of such complexity \cite{Vajenine_2001,Auffermann_2001,Gregoryanz_2004,Young_2006,Crowhurst_2006,Montoya_2007,Chen_2007,
Crowhurst_2008,Wessel_2010,Aydin_2012,Zhao_2014,Wang_2015,Zhang_2016,Bykov_2019,Schneider_2013}. Poly-nitrogen compounds have been considered as potential high-energy density materials, with a remarkable difference of the average bond energy between the single N$-$N (160 kJ mol$^{-1}$), double N$=$N (418 kJ mol$^{-1}$), and triple N$\equiv$N (945 kJ mol$^{-1}$) covalent bonds \cite{Luo_2007}. Moreover, due to the strongly covalent nature of the N$-$N and $M$--N bonding these systems are often considered as candidates for ultra-hard low-compressible materials ({in which} short covalent bonds prevent atomic displacements resulting in exceptional mechanical properties of these solids). Nonetheless, in spite of intensive research, the crystal chemical and physical properties of the binary transition metal nitrides are still poorly understood both experimentally and theoretically to compare, e.g., with oxide materials.

Very recently the high-pressure synthesis and crystal chemical analysis of the binary Fe$-$N system \cite{ Suzuki_1993,Clark_2017,Bykov_2018,Bykov_2018b,Laniel_2018,
Laniel_2018b,Laniel_2022} show the existence of a series different (nitrogen-rich)
compositions, e.g., Fe$_3$N$_2$, FeN, FeN$_2$, and FeN$_4$, stable under pressure \cite{Bykov_2018}. Our particular interest is devoted to the high-pressure (marcasite) phase of FeN$_2$, which contains both the chemical bonding of nitrogen ions (the N-N structural dimers in the lattice structure) \cite{Bykov_2018,Bykov_2018b,Luo_2007,Wessel_2011,Wang_2017} and the effects of strong correlations and magnetic interactions {(spin fluctuations)} of the partially occupied Fe $3d$ states \cite{Mott_1990, Imada_1998}. 

The properties of this model system (FeN$_2$) are also of interest in light of its structural and (partly) electronic similarity to the high-pressure pyrite phase of FeO$_2$ ({with the $Pa\bar{3}$ crystal structure}) \cite{Hu_2016,Nishi_2017}. Recently synthesized under high-pressure and temperature conditions, FeO$_2$ has proven to be stable at the Earth's lower-mantle conditions. Because of this, FeO$_2$ is considered as a geologically important system with an excessive amount of oxygen. Its crystal chemical properties are exceptionally important for understanding of the Earth's lower-mantle properties, and oxygen-hydrogen cycles, have recently been widely debated in the literature \cite{Hu_2016,Nishi_2017,Boulard_2019,Hu_2020,
Koemets_2021,Jang_2017,Streltsov_2017,Lu_2018,Jang_2019,
Liu_2019,Shorikov_2018}.

It was found that FeN$_2$ crystallizes in the trigonal $R\bar{3}m$ structure.
Upon compression above 22 GPa, FeN$_2$ undergoes a structural phase transition to the orthorhombic $Pnnm$ structure as documented using high-resolution powder and single-crystal synchrotron x-ray diffraction \cite{Laniel_2018,Bykov_2018}. The crystal structure consists of chains of edge-sharing FeN$_6$ octahedra aligned along the $c$ axis, which are interconnected through common vertices. Both the $R\bar{3}m$ and $Pnnm$ crystal structures contain structural N$-$N dimers. 
Under pressure of about 59 GPa the nitrogen-nitrogen bond distance in FeN$_2$ is only 1.317 \AA\ \cite{Bykov_2018}, and that is intermediate between the expected bond lengths for the double N$=$N  [N$_2$]$^{2-}$ and single-bonded [N$_2$]$^{4-}$ N$-$N bonds. For example, at ambient conditions  the N$=$N bond length in BaN$_2$ is about 1.23 \AA\ \cite{Wessel_2010}, whereas the calculated single N$-$N bond lengths in PtN$_2$ and OsN$_2$ (with an electronic configuration [N$_2$]$^{4-}$) are $\sim$1.41 and 1.43 \AA\ \cite{Chen_2007,Montoya_2007}, respectively. Note that the effects of electronic correlations and chemical bonding in FeN$_2$ still remain unexplored. This is the main goal of our present study.

In our paper, we explore the electronic structure and magnetic properties of the high-pressure orthorhombic phase of the recently synthesized $Pnnm$ FeN$_2$ using the DFT+dynamical mean-field theory method to study strongly correlated materials \cite{Kotliar_2004,Georges_1996,Kotliar_2006}. Applications of DFT+DMFT have been shown to provide a good quantitative description of the electronic structure and magnetic properties of materials with correlated electrons \cite{Kotliar_2004,Georges_1996,Kotliar_2006}. Using DFT+DMFT it becomes possible to treat on the same footing the spin, charge, orbital, and temperature dependent interactions of the $d$ (or $f$) electrons, as well as to explain a transition from localized to itinerant moment behavior near the {(orbital-selective)} Mott transition (under pressure, doping, or other means) \cite{Kotliar_2004,Georges_1996,Kotliar_2006,Mott_1990, Imada_1998}. We discuss the effects of electron-electron correlations on the electronic structure, magnetic, and crystal chemical properties of this system.

\section{Methods}

We use the DFT+DMFT method to study the spectral properties, local magnetic moments, quasiparticle mass renormalizations of the Fe $3d$ states, and magnetic correlations of the paramagnetic $Pnnm$ phase (PM) of FeN$_2$ under pressure (at about 59 GPa). Moreover, we perform analysis of the crystal chemical properties of FeN$_2$. In our calculations, we use the lattice parameters and atomic positions obtained from the synchrotron x-ray diffraction experiments (space group $Pnnm$ with the lattice parameters $a = 4.4308$ \AA, $b=3.7218$ \AA, and $c=2.4213$ \AA) \cite{Bykov_2018}. 
In particular, in order to quantify the nature of chemical bonding of the N-N dimer states, e.g., the possible formation of the covalent N-N molecular orbitals, 
we perform analysis of the valence electron density plots and determine an electronic configuration of the Fe ion and that of the dinitrogen N-N dimer unit [N$_2$]$^{n-}$. 

In our calculations we employ a fully charge self-consistent implementation of the DFT+DMFT method \cite{Pourovskii_2007,Haule_2007,Aichhorn_2009,Amadon_2012,Park_2014,Hampel_2020} based on the plane-wave pseudopotential formalism within DFT \cite{Leonov_2020}. In DFT we use generalized gradient approximation with the Perdew-Burke-Ernzerhof (PBE) exchange functional as implemented in the Quantum ESPRESSO package with ultrasoft pseudopotentials \cite{Giannozzi_2009}. 
In order to treat the effects of electron correlations in the partially occupied Fe $3d$ shell and charge transfer between the Fe $3d$ and N $2p$ valence states we construct a low-energy {(Kohn-Sham) Hamiltonian $\hat{H}^\mathrm{KS}_{\sigma,mm'}(\mathbf{k})$ using a basis set of atomic-centered Wannier functions for the Fe $3d$ and N $2p$ orbitals \cite{Anisimov_2005,Trimarchi_2008,Korotin_2008}. In DFT+DMFT $\hat{H}^\mathrm{KS}_{\sigma,mm'}(\mathbf{k})$ is supplemented with the on-site Coulomb term for the Fe $3d$ orbitals (in the density-density approximation):
\begin{eqnarray}
\label{eq:hamilt}
\hat{H} = \sum_{\bf{k},\sigma} \hat{H}^{\mathrm{KS}}_{\sigma,mm'}({\bf{k}}) + \frac{1}{2} \sum_{\sigma\sigma',mm'} U_{mm'}^{\sigma\sigma'} \hat{n}_{m\sigma} \hat{n}_{m'\sigma'} - \hat{V}_{\mathrm{DC}}.
\end{eqnarray}
Here, $\hat{n}_{m\sigma}$ is the occupation number operator  (diagonal in the local basis set) with spin $\sigma$ and orbital indices $m$. $U_{mm'}^{\sigma\sigma'}$ denotes the reduced density-density form of the four-index Coulomb interaction matrix: $U_{mm'}^{\sigma\overline{\sigma}}=U_{mm'mm'}$ and $U_{mm'}^{\sigma\sigma}=U_{mm'mm'}-U_{mm'm'm}$. The latter is expressed in terms of the Slater integrals $F^0$, $F^2$, and $F^4$. For the $3d$ electrons these parameters are related to the Coulomb and Hund's coupling as $U=F^0$, $J=(F^2+F^4)/14$, and $F^2/F^4=0.625$. $\hat{V}_\mathrm{DC}$ is the double-counting correction to account for the electronic interactions described within DFT. Here, we use the fully localized double-counting correction evaluated from the self-consistently determined local occupations. Our results obtained using DFT+DMFT with the around mean-field double counting (AMF) are shown in Supplemental Material (SM) \cite{suppl}.}

We use the continuous-time hybridization expansion quantum Monte Carlo method (CT-QMC) to treat the many-body and strong correlations effects in the Fe $3d$ shell \cite{Gull_2011}. In our calculations we employ the CT-QMC segment algorithm {neglecting for simplicity} the pair and spin-flip hopping terms of the Hund's exchange {( contributions of which are typically small) \cite{rot_inv}}. The effects of electron correlations in the Fe $3d$ orbitals are treated by using the on-site Hubbard $U = 6$ eV and Hund's exchange $J = 0.89$ eV, taken in accordance with previous calculations \cite{Leonov_2015,Leonov_2016,Leonov_2020,Leonov_2019,Layek_2022,Greenberg_2023,
Greenberg_2018,Ohta_2012,Kunes_2009}. {The N $2p$ states are considered as non-correlated and are treated on the DFT level within the fully charge self-consistent DFT+DMFT approach \cite{Leonov_2015,Leonov_2016,Leonov_2020,Leonov_2019}.}
{We also check different sets of the Hubbard $U$ and Hund's exchange $J$ values for the Fe $3d$ states. Moreover, to treat correlation effects on the partially occupied N $2p$ orbitals we perform the DFT+DMFT calculations with $U=4$ and $J=0.5$ eV for the N $2p$ electrons (for the Fe sites we use $U = 6$ eV and $J = 0.89$ eV).} 
In our calculations we neglect the spin-orbit coupling effects. In order to compute the {\bf k}-resolved spectra we use Pad\'e approximants to perform analytic continuation of the self-energy results on the real energy axis. {In addition, we employ the maximum entropy method for analytic continuation of the Green's functions data \cite{Sandvik_1998}.}
Using DFT+DMFT we study the electronic structure and magnetic properties of the high-pressure PM FeN$_2$ at a temperature $T = 290$ K.

\section{RESULTS AND DISCUSSION}

\subsection{Electronic structure}

\begin{figure}[tbp!]
\centerline{\includegraphics[width=0.5\textwidth,clip=true]{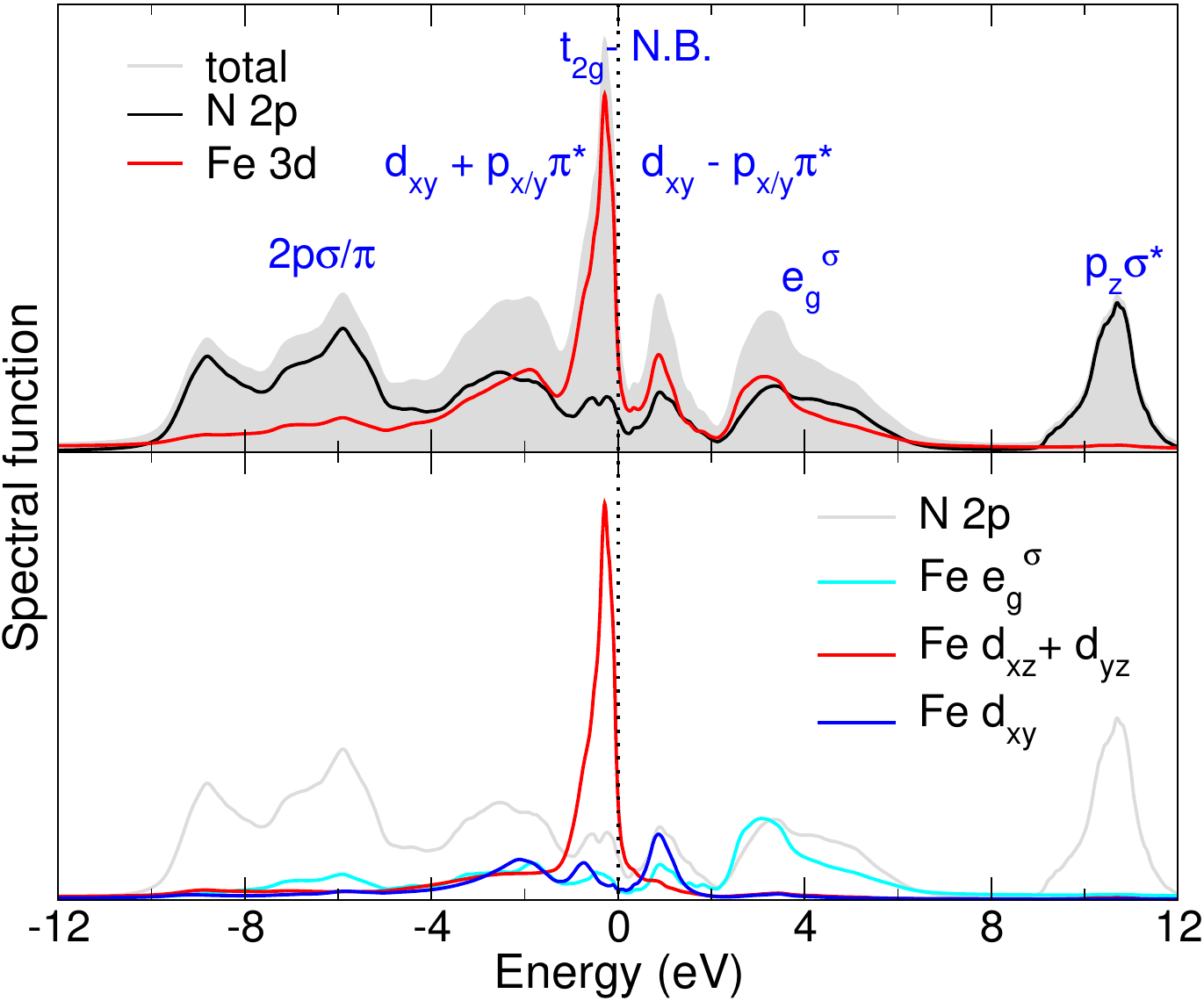}}
\caption{
Our results for the orbital-dependent spectral functions of paramagnetic FeN$_2$ as obtained by DFT+DMFT at $T=290$ K. The partial Fe $3d$ ($d_{xz}+d_{yz}$, $d_{xy}$, and $e_g^\sigma$) and N $2p$ (bonding: $2p$ $\sigma$ and $2p$ $\pi$, and antibonding: $2p_{x/y}$ $\pi^*$ and $2p_z$ ${\sigma^*}$) orbital contributions are shown. Note the empty antibonding N $2p_z$ ${\sigma^*}$ states located at about 10 eV above the Fermi level. 
{While the Fe $xz$/$yz$ and $x^2-y^2$/$3z^2-r^2$ orbitals are not degenerate in the orthorhombic 
phase of FeN$_2$, the differences between them are small 
and for the sake of clarity only the total contributions are shown.}}
\label{Fig_1}
\end{figure}

\begin{figure}[tbp!]
\centerline{\includegraphics[width=0.5\textwidth,clip=true]{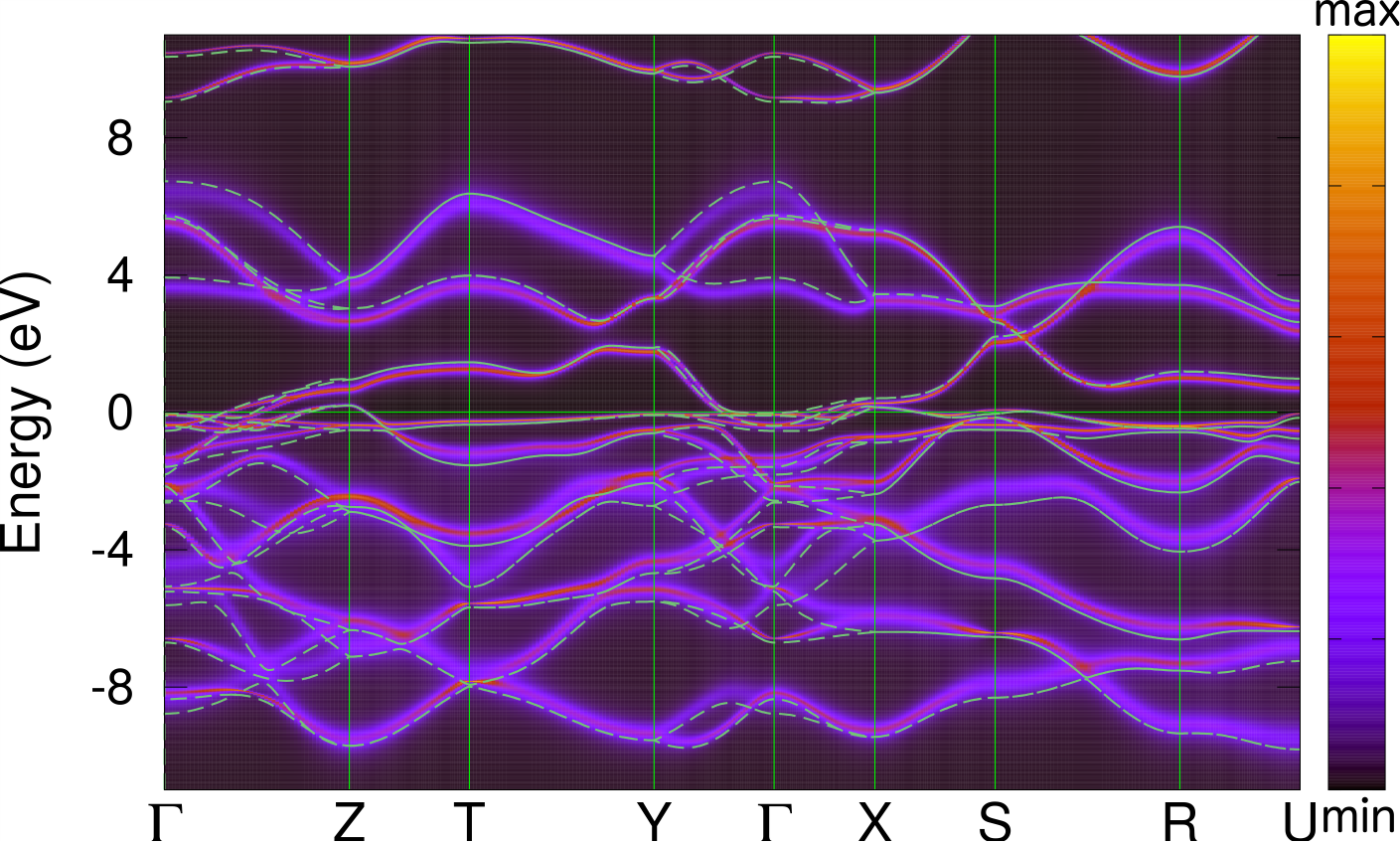}}
\caption{{\bf k}-resolved spectral functions calculated by DFT+DMFT at $T = 290$ K for the orthorhombic $Pnnm$ crystal 
structure of FeN$_2$. {Our nm-DFT (PBE) results are shown with green broken green lines.}
}
\label{Fig_2}
\end{figure}

In Fig.~\ref{Fig_1} we display our results for the orbital dependent Fe $3d$ and N $2p$ spectral functions of FeN$_2$ calculated by DFT+DMFT at 290 K (with the crystal structure parameters taken at about 59 GPa from Ref.~\onlinecite{Bykov_2018}). Our results for the {\bf k}-resolved spectral functions are shown in Fig.~\ref{Fig_2}. 
{ Our results for the spectral function agree well with those obtained by the nonmagnetic DFT (nm-DFT), suggestive of small quasiparticle mass renormalizations and, hence, weak correlations in FeN$_2$ under pressure (see SM~\cite{suppl})}.
We obtain metal with the partially occupied Fe $t_{2g}$ states located near the Fermi level which form {coherent electronic states} between -4 and 2 eV near the Fermi level. The Fe $e_g^\sigma$ {($d_{x^2-y^2}$ and $d_{3z^2-r^2}$ orbitals)} states are empty and appear between 2 and 6 eV above the $E_F$. It leads to a relatively large crystal-field splitting between the $t_{2g}$-$e_g$ orbitals of about 4 eV, which in turn is suggestive of the low-spin state (LS) of Fe ions (i.e., crystal-field effects dominate over the Hund's exchange coupling $J$). In contrast to this, our DFT+DMFT results yield a relatively large (for the LS Fe$^{2+}$ ion) instantaneous local magnetic moment value $\sqrt{\langle \hat{m}^2_z \rangle}$ of 1.56 $\mu_\mathrm{B}$ per Fe ion, instead of the non-magnetic $t_{2g}$ LS configuration of the Fe$^{2+}$ ions as one can expect under high pressure. 
{It is worth noting that the nm-DFT and DFT+DMFT calculations give a relatively large bandwidth of the N $2p$ orbitals, of about 16 eV (excluding the nonbonding N $p_z\sigma$ 
states located at about 10 eV above the Fermi level), implying weak correlations in the N $2p$ bands~\cite{note_u_on_p}}.

It is also interesting to note that the crystal structure of FeN$_2$ determined from single-crystal x-ray diffraction experiments at about 59 GPa (space group $Pnnm$) contains the N-N structural dimers with an interatomic distance of $\sim$1.317 \AA \cite{Bykov_2018}. This implies the possible formation of the N-N molecular bonding state in FeN$_2$. In the N$_2$ molecule, the N-N bond length distance is about 1.1 \AA, while a charge transfer in FeN$_2$ yields [N$_2$]$^{n-}$ (with $n>0$) with larger N-N covalent bond length (a charge transfer between Fe and N ions controls the length of the N-N bond). In agreement with this (a picture of the covalent N-N bonding) we observe a large bonding-antibonding splitting of the N $2p$ states. In fact, the bonding N $2p$ $\sigma$ and $\pi$ states are fully occupied and appear deep below the Fermi level between about -10 and -4 eV. In contrast to this the antibonding N $p_z$ $\sigma^*$ states are empty (like in the N$_2$ molecule) and form bands located at about 10 eV above the $E_F$ (see Fig.~\ref{Fig_1}). 

Moreover, the partially occupied $\pi$-type antibonding $[\mathrm{N}_2]^{n-}$ $2p_x$ $\pi^*$ and $2p_y$ $\pi^*$ states appear near the Fermi level (at the place where the Fe $t_{2g}$ states lie), strongly hybridizing with the partially occupied Fe $t_{2g}$ states. This leads to the appearance of a complex bonding structure near the Fermi level. In particular, our results suggest the formation of a quasiparticle peak at about -0.3 eV primarily originating from the non-bonding Fe $t_{2g}$ states (of the Fe $xz$ and $yz$ orbital character in the local coordinate frame
with the $z$-axis along the shortest Fe-N bond). The non-bonding Fe $xz$ and $yz$ orbitals form a weakly dispersive {band} of about 3 eV bandwidth with a peak at about -0.3 eV below the Fermi level (it is labeled as Fe $t_{2g}$-N.B. in Fig.~\ref{Fig_1}). 

At the same time, the Fe $xy$ and N $p_{x/y}$ $\pi^*$ orbitals form bonding and antibonding combinations, the $d_{xy}+p_{x/y} \pi^*$ and $d_{xy}-p_{x/y} \pi^*$ molecular orbital states, respectively, which are partially occupied and are located between -4 and 2 eV near the $E_F$. We find that the bonding $d_{xy}+p_{x/y}~\pi^*$ orbitals are fully occupied and appear at -2.1 eV below $E_F$, while the antibonding states are empty and are at about 0.9 eV above the Fermi level. It gives a relatively large bonding-antibonding splitting of $\sim$3 eV between the $d_{xy}+p_{x/y} \pi^*$ and $d_{xy}-p_{x/y} \pi^*$ states. Overall, the Fe $t_{2g}$ spectrum exhibits a three peak structure which is a typical characteristic feature of strongly correlated metals (with a quasiparticle peak and the lower and upper Hubbard subbands). However, in FeN$_2$ this behavior has different origins and is associated with complex bonding of the N $2p$ and Fe $3d$ states due to the formation of bonding, antibonding, and non-bonding $p$-$d$ orbital states with a large bonding-antibonding splitting.

\subsection{N-N bonding}

\begin{figure*}[tbp!]
\centerline{\includegraphics[width=\textwidth,clip=true]{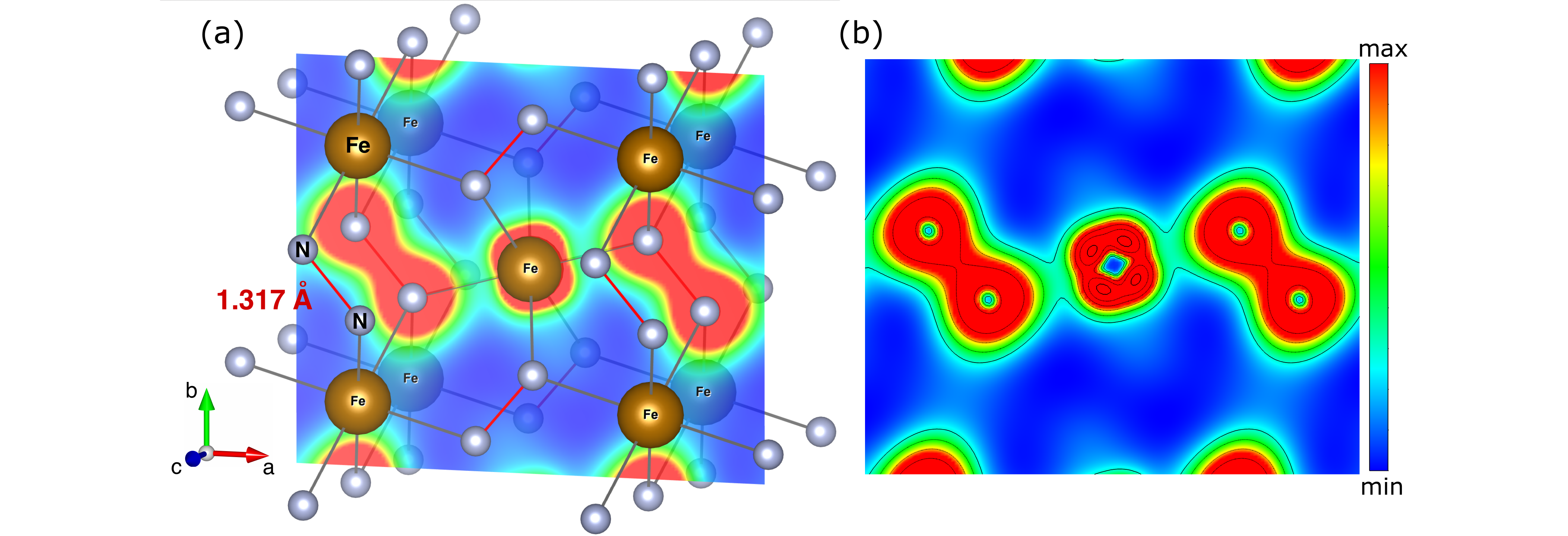}}
\caption{Crystal structure and valence electron density plots obtained by DFT+DMFT at  $T=290$ K for the $Pnnm$ crystal structure of FeN$_2$ (under pressure about 59 GPa). The N-N dimer with a bond length distance $\sim$1.317 \AA\ in (a) is shown in red. Max stands for 25\% of maximum of charge density $\rho({\bf r})$. {Our results of the DFT+DMFT (AMF) calculations are given in SM~\cite{suppl}.}
}
\label{Fig_3}
\end{figure*}

We further quantify this point by computing the valence electron density distribution (a projection of the total charge density on the $ab$ plane cutting the N-N bond and the Fe ions). In Fig.~\ref{Fig_3} we plot our results for the valence electron density distribution obtained by DFT+DMFT for the $Pnnm$ crystal structure of FeN$_2$ under pressure of about 59 GPa. We note that the N-N structural dimers clearly show the formation of a strongly covalent N-N bond with a large charge density at the N-N contact, which is about 36\% of a maximal electron density value. For comparison, in the pyrite-type MgO$_2$ peroxide with a covalent bond [O$_2$]$^{2-}$ this value is $\sim$21\% \cite{Koemets_2021}. It is interesting that this differs qualitatively from the previously obtained DFT+DMFT results for the pyrite-structured FeO$_2$ in which this value was significantly smaller, only about 5\%, and hence the authors claim the absence of covalent oxygen-oxygen bonding in FeO$_2$ (with a 1.5- valence state of oxygen) \cite{Koemets_2021}. Our results therefore support the formation of a strong covalent N-N bond in FeN$_2$ under pressure, in agreement with experimental data \cite{Bykov_2018}.

While the electronic structure of FeN$_2$ near the Fermi level strongly resembles that of FeO$_2$, the main distinction is that the antibonding N $2p$ $\pi^*$ states appear at the $E_F$ (in FeO$_2$ these are the O $2p$ $\sigma^*$). Furthermore, in FeN$_2$ a strongly covalent N-N bonding results in a large bonding-antibonding splitting of the $2p$ states, with the N $2p_z$ $\sigma^*$ states located well above $E_F$, at $\sim$10 eV.

As we noted the calculated spectral functions of PM FeN$_2$ show the empty antibonding $d_{xy}-p_{x/y} \pi^*$ states located above $E_F$, implying that the Fe $t_{2g}$ states are partially occupied. Our results therefore suggest that instead of a formal valence state of iron Fe$^{2+}$ and the double-bonded N$=$N dimer [N$_2$]$^{2-}$ in FeN$_2$, we deal with the ferric Fe$^{3+}$ ions. Interestingly, this 
result is in accordance with the Fe$^{3+}$ valence state previously proposed in the cubic pyrite-structured FeO$_2$ \cite{Koemets_2021,Streltsov_2017}. This result is also consistent with our estimate of the total charge at the Fe site inside the atomic sphere with a radius $\sim$0.78 \AA\ (an integral of the charge density around the Fe site with a given ionic radius) which gives 4.86 electrons, i.e., a nearly Fe$^{3+}$ $3d^5$ state. A similar estimate for a radius 0.9 \AA\ gives 5.61. 
{Note that the DFT+DMFT (AMF) calculations give 4.88 and 5.57 electrons, respectively. Overall, we found that the total charge at the Fe site is robust, varying within $\sim1$\% upon changing the Hubbard $U$ and Hund's coupling $J$ within 4-6 eV and 0.45-0.89 eV, respectively, for the Fe sites, as well as upon applying the Coulomb correlations on the N $2p$ orbitals.}
Moreover, the Fe$^{3+}$ ion in the low-spin state is consistent with a relatively large (instantaneous) local moment value of about 1.56 $\mu_\mathrm{B}$ per Fe ion. In accordance with this, our analysis of the weights of different spin-state configurations of the Fe $3d$ electrons (fluctuating between various atomic configurations within DMFT) shows a predominant low-spin configuration with a weight of 66\%, with a notable admixture of the intermediate-spin states (17\%).

\begin{figure}[tbp!]
\centerline{\includegraphics[width=0.5\textwidth,clip=true]{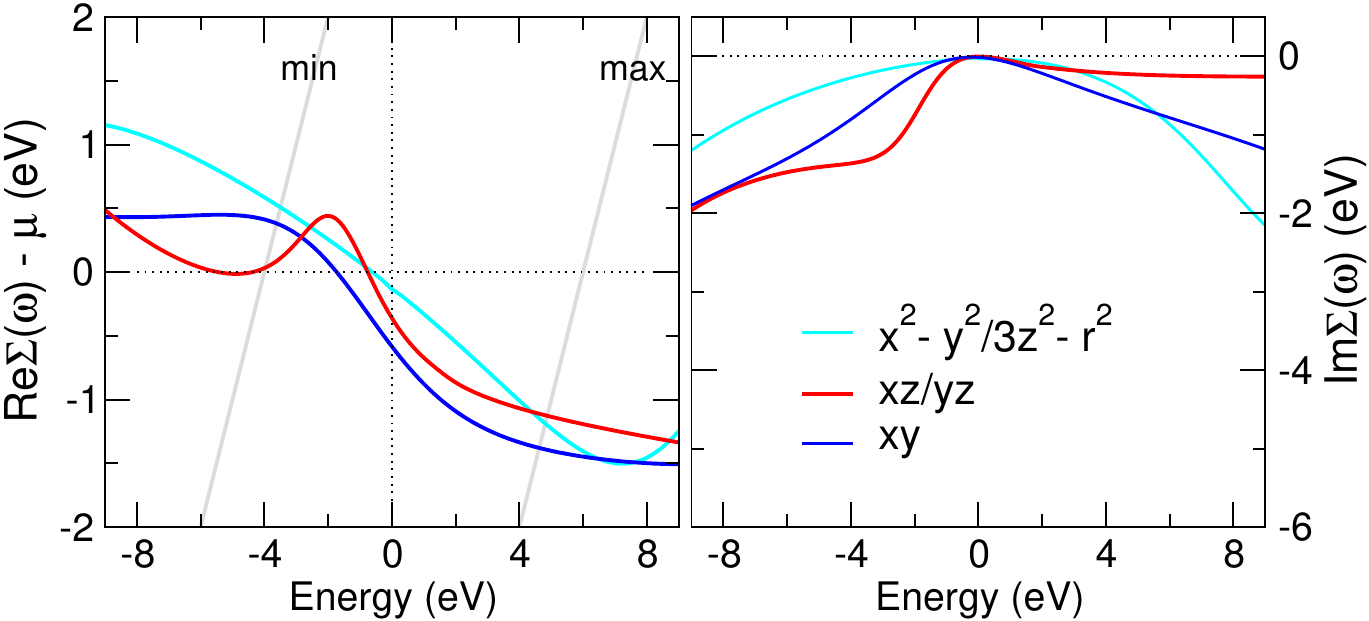}}
\caption{Our results for the orbital-dependent Fe $3d$ self-energies obtained by DFT+DMFT at $T=290$ K analytically continued on the real energy axis $\Sigma(\omega)$ using Pad\'e approximants. {min (max) stands for the minimal (maximal) energy range of the Fe $3d$ bands obtained within nm-DFT. Intersections with the real part of the self-energies roughly give the poles of the {\bf k}-resolved Green's function, as determined by $\mathrm{det}[\hat{H}^\mathrm{DFT}({\bf k})+\mathrm{Re}\hat{\Sigma}(\omega)  - \omega]=0$. The non-trivial solutions of this equation show the position of the lower and upper Hubbard subbands \cite{Poteryaev_2007}. Our analysis show the absence of the Hubbard subbands in the electronic structure of FeN$_2$, implying a weakly correlated (itinerant) behavior of the Fe $3d$ electrons.}
}
\label{Fig_4}
\end{figure}

We note that an unexpected valence state of Fe$^{3+}$ ions implies that the N-N dimer states adopt a formal electronic configuration [N$_2$]$^{3-}$. Most importantly, this result agrees well with our analysis of the N-N bond lengths. In fact, the N-N bond length of about 1.317 \AA\ in FeN$_2$ is found to be intermediate to those expected for the double and single-bonded N-N  dimers \cite{Bykov_2018}. In particular, at ambient conditions the N$=$N bond lengths in BaN$_2$ (the [N$_2$]$^{2-}$ state) is about 1.23 \AA\ \cite{Wessel_2010}, whereas the calculated N$-$N bond lengths in [N$_2$]$^{4-}$ in PtN$_2$ and OsN$_2$ are significantly higher, about 1.41 and 1.43 \AA, respectively \cite{Chen_2007,Montoya_2007}. Comparing the N-N bond length in FeN$_2$ (1.317 \AA) with previous estimates of the characteristic  bond lengths, e.g., with the N-N bond lengths typical for the N$_2$ ($\sim$1.1 \AA), [N$_2$]$^{2-}$  (1.23-1.27 \AA), and  [N$_2$]$^{3-}$ (1.30-1.34 \AA) molecular orbital complexes \cite{Laniel_2022}, we propose that the N-N dimers adopt a valence state close to [N$_2$]$^{3-}$. While previous reports suggest the possible formation of the [N$_2$]$^{2-} + \bar{e}$ electride-like state \cite{Dye_2009,Liu_2020} in the Fe-N systems \cite{Wessel_2010,Schneider_2013}, our analysis of the electronic structure gives no evidence for this state. In contrast to this, our results suggest strong bonding of the N $2p_{x/y}$ $\pi^*$ and Fe $d_{xy}$ states near the Fermi level, which in turn stabilizes the [N$_2$]$^{3-}$ radical ions.

\subsection{Quasiparticle mass renormalizations and magnetic correlations}

\begin{figure}[tbp!]
\centerline{\includegraphics[width=0.5\textwidth,clip=true]{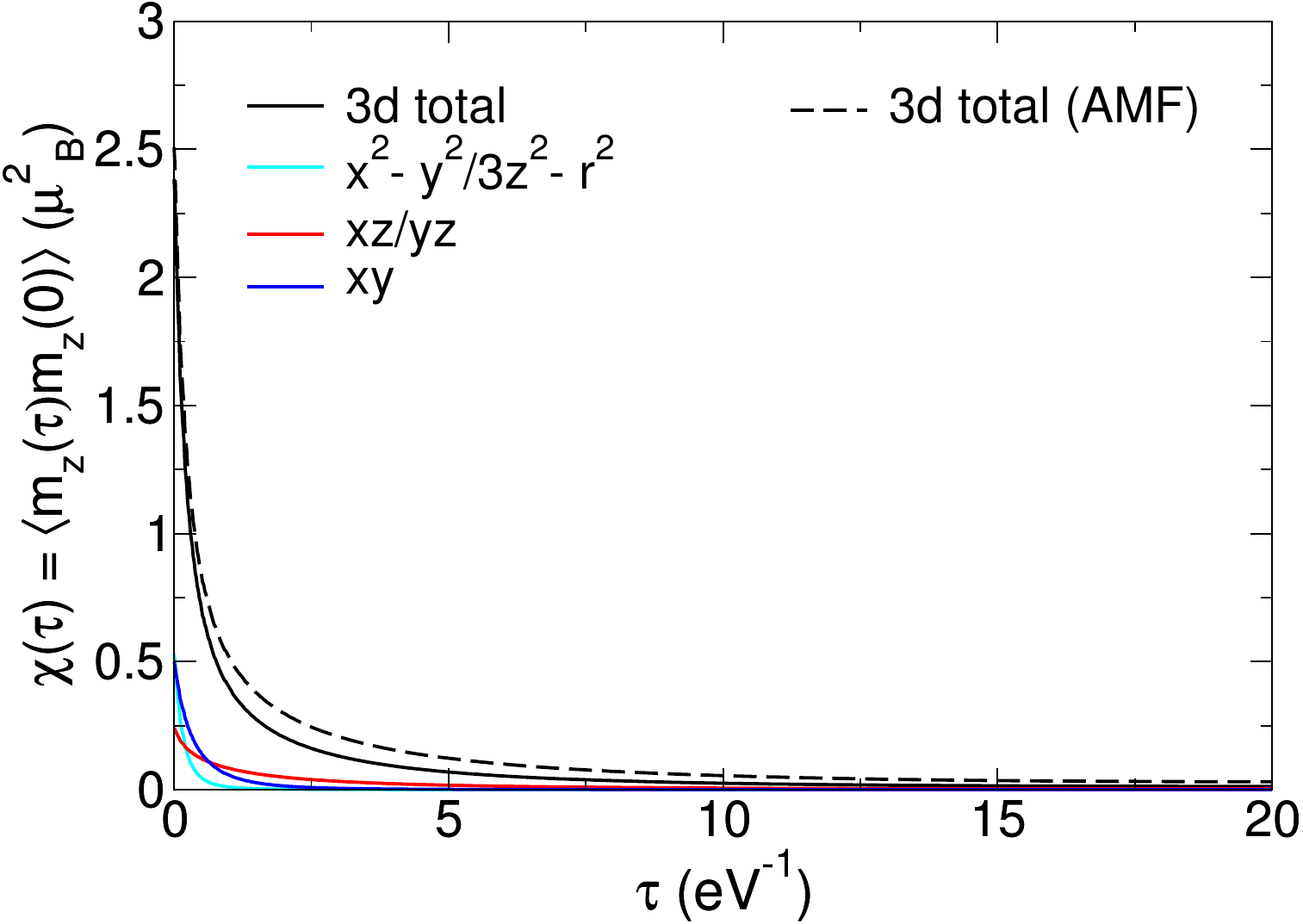}}
\caption{Orbital-resolved local spin correlation functions $\chi(\tau) = \langle \hat{m}_z(\tau) \hat{m}_z(0) \rangle$ as a function of the imaginary time $\tau$ for the Fe $3d$ orbitals calculated by DFT+DMFT for FeN$_2$ at $T = 290$ K. {Our results obtained by DFT+DMFT with the around mean-field double counting are depicted as AMF.}
}
\label{Fig_5}
\end{figure}

Next, we discuss the effects of electron correlations on the electronic structure and magnetic properties of FeN$_2$ under pressure. Our results for the Fe $3d$ self-energies analytically continued on the real energy axis $\Sigma(\omega)$ using Pad\'e approximants are shown in Fig.~\ref{Fig_4}.  Our results for the Fe $3d$ self-energies on the Matsubara contour suggests a typical Fermi liquid-like behavior, with highly coherent spectral weights and weak quasiparticle damping of the Fe $3d$ states. Thus, $\mathrm{Im}[\Sigma(i\omega_n)]\sim 0.03$ and 0.05 eV for the Fe $t_{2g}$ and $e_g^\sigma$ orbitals at the first Matsubara frequency, respectively, at $T = 290$K. Moreover, our DFT+DMFT calculations reveal weak renormalizations of the Fe $3d$ states $m^*/m = [1 - \partial \mathrm{Im}[\Sigma(i\omega)]/ \partial i\omega]|_{i\omega\rightarrow 0}$, $\sim$1.2$-$1.4, with no formation of the lower and upper Hubbard subbands in the Fe $3d$ spectral function.
{In addition, $\Sigma(\omega)$ gives no non-trivial poles in the {\bf k}-resolved Green's function  (see Fig.~\ref{Fig_4}) defined as $\mathrm{det}[\hat{H}^\mathrm{DFT}({\bf k})+\mathrm{Re}\hat{\Sigma}(\omega)  - \omega]=0$ \cite{Poteryaev_2007}.}
Our analysis suggests weak (orbital-dependent) correlation effects with $m^*/m \sim 1.4$, 1.3, and 1.2 for the Fe $xz$/$yz$, $xy$, and $e_g^\sigma$ states, respectively.
{We note that the quasiparticle band renormalizations depend weakly on the choice of the double counting scheme within DFT+DMFT. Using DFT+DMFT (AMF) we obtain $m^*/m \sim 1.5$, 1.4 and 1.2 for the Fe $xz$/$yz$, $xy$, and $e_g^\sigma$ orbitals, respectively. Moreover, our results for $m^*/m$ show a rather weak dependence upon variations of the Hubbard $U$ and Hund's exchange $J$ parameters. In particular, for $U=4$ eV and $J=0.45$ eV we obtain  $m^*/m \sim 1.3$, 1.2, and 1.1 for the Fe $xz$/$yz$, $xy$, and $e_g^\sigma$ orbitals, respectively, while for $U=6$ eV and $J=0.45$ eV these values are 1.4, 1.3, and 1.2.}
This implies a weakly correlated (itinerant magnetic moment) behavior of the Fe $3d$ states in FeN$_2$ under pressure. In accordance with this our analysis of the orbital-dependent local spin susceptibility $\chi(\tau)$ is evaluated within DFT+DMFT (see Fig.~\ref{Fig_5}). Our calculations show a fast decay of $\chi(\tau)$ for the Fe $3d$ states to about 0.01 $\mu_\mathrm{B}^2$ at $\tau=\beta/2$. It is consistent with our estimate of the fluctuating moment evaluated as $M_\mathrm{loc} \equiv [k_\mathrm{B}T \int \chi(\tau) d\tau]^{1/2}$ (where $\chi(\tau) \equiv \langle \hat{m}_z(\tau)\hat{m}_z(0) \rangle $ is the local spin correlation function), of about 0.3 $\mu_\mathrm{B}$, which differs significantly from the instantaneous moments 1.56 $\mu_\mathrm{B}$ [$\langle \hat{m}_z^2\rangle \equiv \chi(\tau=0)$].

\begin{figure}[tbp!]
\centerline{\includegraphics[width=0.5\textwidth,clip=true]{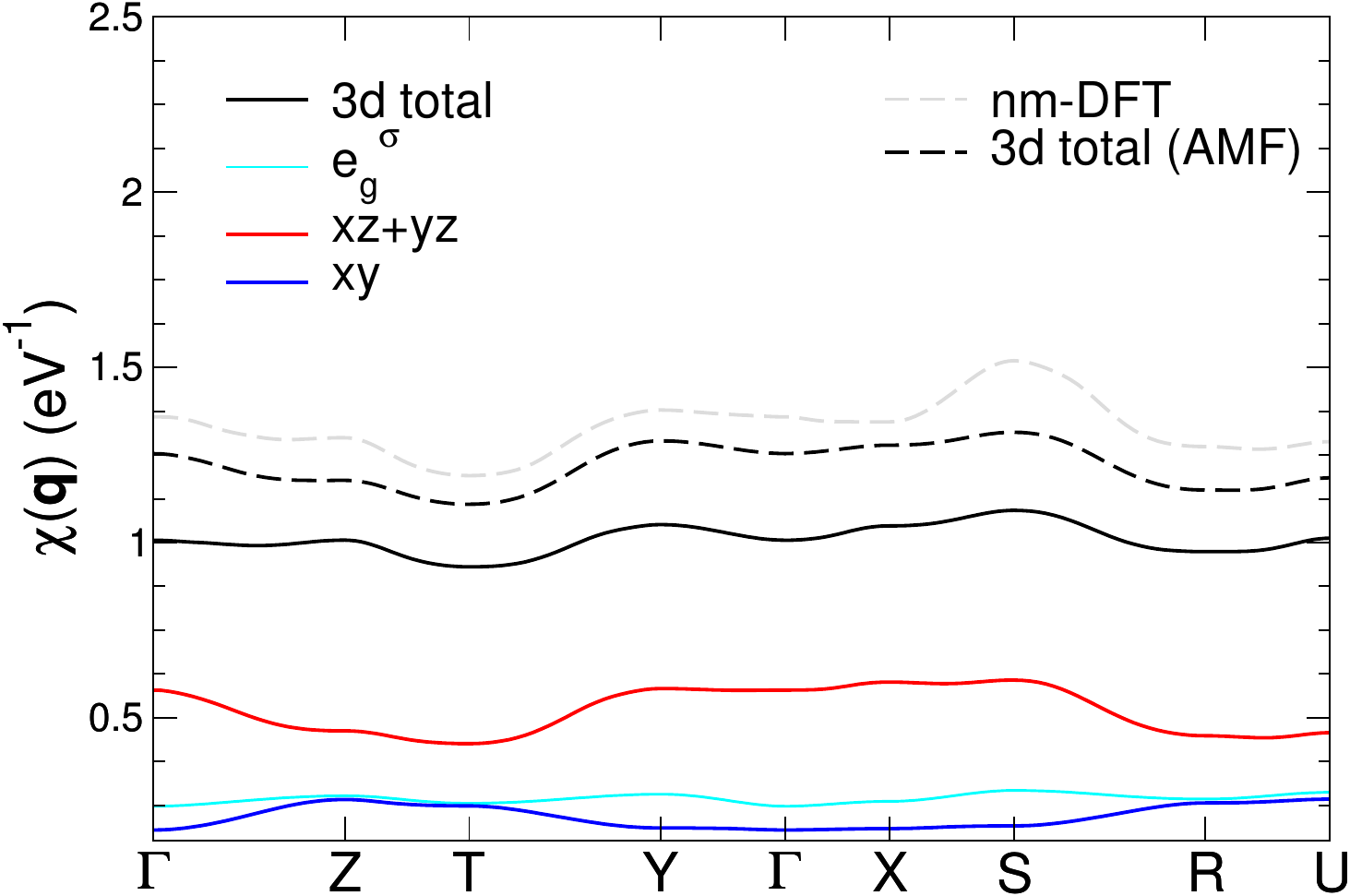}}
\caption{Orbitally resolved static spin susceptibility $\chi({\bf q})$ of Fe ions calculated by DFT+DMFT for the high-pressure orthorhombic FeN$_2$ at $T=290$ K. { For comparison we show the Fe $3d$ total $\chi({\bf q})$ results obtained by the nm-DFT and DFT+DMFT (AMF) calculations. $\chi({\bf q})$ is evaluated as $\chi({\bf q})=-k_BT\mathrm{Tr} \Sigma_{{\bf k}, i\omega_n}G_{\bf k}(i\omega_n)G_{{\bf k}+{\bf q}}(i\omega_n)$, where $G_{\bf k}(i\omega_n)$ are the matrix elements of the local Green's function for the Fe $3d$ states.}}
\label{Fig_6}
\end{figure}

Finally, we study the effects of long-range magnetic correlations characterized by an ordering wave vector driven by the nesting wave vector of the Fermi surface of FeN$_2$. {For PM FeN$_2$} we determine the momentum-dependent static magnetic susceptibility $\chi({\bf q})$ in the particle-hole bubble approximation within DFT+DMFT as $\chi({\bf q})=-k_BT\mathrm{Tr} \Sigma_{{\bf k}, i\omega_n}G_{\bf k}(i\omega_n)G_{{\bf k}+{\bf q}}(i\omega_n)$, where $G_{\bf k}(i\omega_n)$ {are the matrix elements of the local \emph{interacting} Green's function} 
for the Fe $3d$ states evaluated on the Matsubara contour $i\omega_n$. {To compute $\chi({\bf q})$ we take a trace over the orbital indices, denoted by Tr.} Our results for $\chi({\bf q})$ are shown in Fig.~\ref{Fig_6}. $\chi({\bf q})$ is seen to show multiple well-defined maxima, suggesting a complex interplay between different spin density wave states (with commensurate wave vectors) in FeN$_2$ under pressure. We note that the most pronounced instability is associated 
with a wave vector S $(\frac{1}{2}~\frac{1}{2}~0)$, competing with two minor instabilities at the X $(\frac{1}{2}~0~0)$ and Y  $(0~\frac{1}{2}~0)$  points of the Brillouin zone. 
{
Moreover, our results for $\chi({\bf q})$ remain qualitatively the same upon variations of the $U$ and $J$ values, with the most pronounced instability at S, competing with two 
minor instabilities at the X and Y points.}
This implies the possible emergence of long-range multiple (or intertwined) spin density wave ordering states in FeN$_2$ at low pressures. This may lead to the emergence of antiferromagnetic spin fluctuations in FeN$_2$ which makes this system a possible candidate for spin-fluctuation mediated superconductivity. This subject deserves further detailed theoretical and experimental research.

\section{Conclusions}

In conclusion, using the DFT+DMFT method we computed the electronic structure and magnetic properties of PM FeN$_2$ under high pressure. Our analysis of the electronic states in FeN$_2$ shows complex crystal-chemical behavior characterized by the formation of a strong covalent N-N bond, in agreement with experimental data \cite{Bykov_2018}. Our results show the formations of an unexpected valence state of iron Fe$^{3+}$ (i.e., Fe ions are ferric and low-spin paramagnetic), with the N-N dimer states adopting an electronic configuration [N$_2$]$^{3-}$. This result agrees with previous estimates of the characteristic bond lengths for the [N$_2$]$^{n-}$ molecular orbital complexes for different nitrogen-based systems. Our results suggest weak (orbital-dependent) correlation effects in FeN$_2$, which are complicated by the possible emergence of multiple (intertwined) spin density wave states on a microscopic level. This suggests the importance of antiferromagnetic ordering and spin fluctuations to explain the properties of FeN$_2$ under pressure. {Overall, this topic deserves further detailed experimental and theoretical analysis. In this respect applications of perturbative techniques such as the $GW$ method might be reasonable to treat the electronic correlations without a separation of the Fe $3d$ and N $2p$ orbitals into correlated and uncorrelated subspaces \cite{Hedin_1965,Rohlfing_2000,Aryasetiawan_1998,Onida_2002,Sun_2002,Schilfgaarde_2006,
Klimes_2014,Zhu_2021}. We leave this challenging problem for the future.}


\begin{acknowledgments}
We acknowledge the support of the Ministry of Science and Higher Education of the Russian Federation, project No. 122021000038-7 (theme ``Quantum'').
\end{acknowledgments}

\end{document}